\documentclass{aastex62}

\graphicspath{{./}{figures/}}

\received{January 1, 2018}
\revised{January 7, 2018}
\accepted{\today}
\submitjournal{ApJ}

\shorttitle{OJ~287 accretion disk}
\shortauthors{Valtonen et al.}

\begin{document}

\title{Accretion Disk Parameters determined from the great 2015 flare of OJ~287}

\correspondingauthor{Mauri Valtonen}
\email{mvaltonen2001@yahoo.com}

\author{Mauri J. Valtonen}
\affiliation{FINCA, University of Turku, Turku, Finland}
\affiliation{Department of Physics and Astronomy, University of Turku, Turku, Finland}

\author{Staszek Zola}
\affiliation{Astronomical Observatory, Jagiellonian University, ul. Orla 171, Cracow, Poland}
\affiliation{Pedagogical University of Cracow, Mt. Suhora Astronomical Observatory, Cracow, Poland}

\author{Pauli Pihajoki}
\affiliation{Department of Physics, University of Helsinki, Helsinki, Finland}

\author{Sissi Enestam}
\affiliation{Mets\"ahovi Radio Observatory, Aalto University, Espoo, Finland}

\author{Harry J. Lehto}
\affiliation{Department of Physics and Astronomy, University of Turku, Turku, Finland}

\author{Lankeswar Dey}
\affiliation{Department of Astronomy and Astrophysics, Tata Institute of Fundamental Research, Mumbai, India}

\author{Achamveedu Gopakumar}
\affiliation{Department of Astronomy and Astrophysics, Tata Institute of Fundamental Research, Mumbai, India}

\author{Marek Drozdz}
\affiliation{Pedagogical University of Cracow, Mt. Suhora Astronomical Observatory, Cracow, Poland}

\author{Waldemar Ogloza}
\affiliation{Pedagogical University of Cracow, Mt. Suhora Astronomical Observatory, Cracow, Poland}

\author{Michal Zejmo}
\affiliation{Janusz Gil Institute of Astronomy, University of Zielona Góra, Prof. Szafrana 2, PL-65-516 Zielona Góra,Poland}

\author{Alok C. Gupta}
\affiliation{Aryabhatta Research Institute of Observational Sciences (ARIES), Nainital,  India}

\author{Tapio Pursimo}
\affiliation{Nordic Optical Telescope, Apartado 474, E-38700 Santa Cruz de La Palma, Spain}

\author{Stefano Ciprini}
\affiliation{Agenzia Spaziale Italiana (ASI) Science Data Center, I-00133 Roma, Italy}
\affiliation{Istituto Nazionale di Fisica Nucleare, Sezione di Perugia, I-06123, Perugia, Italy}

\author{Mark Kidger}
\affiliation{Herschel Science Centre, ESAC, European Space Agency, 28691 Villanueva de la Cañada, Madrid, Spain}

\author{Kari Nilsson}
\affiliation{FINCA, University of Turku, Turku, Finland}

\author{Andrei Berdyugin}
\affiliation{Department of Physics and Astronomy, University of Turku, Turku, Finland}

\author{Vilppu Piirola}
\affiliation{FINCA, University of Turku, Turku, Finland}
\affiliation{Department of Physics and Astronomy, University of Turku, Turku, Finland}

\author{Helen Jermak}
\affiliation{Astrophysics Research Institute, Liverpool John Moores University, IC2 Liverpool Science Park, Liverpool L3 5RF, UK}

\author{Rene Hudec}
\affiliation{Astronomical Institute, The Czech Academy of Sciences, 25165 Ondřejov, Czech Republic}
\affiliation{Czech Technical University in Prague, Faculty of Electrical Engineering, Prague, Czech Republic}
and 

\author{Seppo Laine}
\affiliation{IPAC, Mail Code 314-6, Caltech, 1200 E. California Boulevard, Pasadena, CA 91125, USA}



\begin{abstract}

In the binary black hole model of OJ~287 the secondary black hole orbits a much more massive primary, and impacts on the primary accretion disk at predictable times. We update the parameters of the disk, the viscosity $\alpha$ and the mass accretion rate $\dot m$. We find $\alpha=0.26\pm0.1$ and $\dot m=0.08\pm0.04$ in Eddington units. The former value is consistent with \citet{C81} and the latter with \citet{M11}. Predictions are made for the 2019 July 30 superflare in OJ~287. We expect that it will take place simultaneously at the Spitzer infrared channels as well as in the optical and that therefore the timing of the flare in optical can be accurately determined from Spitzer observations. We also discuss in detail the light curve of the 2015 flare, and find that the radiating volume has regions where bremsstrahlung dominates as well as regions that radiate primarily in synchrotron radiation. The former region produces the unpolarised first flare while the latter region gives rise to a highly polarized second flare.

\end{abstract}


\section{Introduction} \label{sec:intro}

OJ~287 is a unique blazar which shows large thermal flares at predictable times. A roughly 12 year cycle of flares was noticed during the Tuorla Observatory monitoring program by \cite{val88} and \cite{S88}. The optical light curve of OJ~287 goes all the way back to the year 1888. The early part of this data set exists due to the proximity of OJ~287 to the ecliptic which is the reason why it was unintentionally photographed often in the past, providing us with such a long light curve. It may be divided in the historical part, pre-1970, and the modern part, post-1970, separated by the time when OJ~287 was recognized as an interesting extragalactic object. Initially, the historical part consisted only of about 200 points which are often displayed even today (see e.g. \cite{britzen18}). However, an extensive study of the plate archives by one of the authors (R.H.) has increased the number of historical light curve points by an order magnitude, including a dense network of upper limits, which demonstrate an interesting pattern of 24 major flares over the 130 yr time span \citep{D18}.

\citet{LV96} found that the pattern of major flares is generated by a simple mathematical rule which may be called the Keplerian sequence. Consider a point moving in a Keplerian elliptical orbit, and let the major axis of the ellipse rotate forward at the same time by a constant rate. The sequence of times when the point crosses a fixed line in the orbital plane, drawn through the Keplerian focal point of the ellipse, forms the Keplerian sequence. For every value of eccentricity and rotation rate of the ellipse, a different sequence is created. If we choose  the eccentricity and the precession rate to be $e\sim 0.7$ and $\Delta \phi\sim 39$ degrees per orbit, respectively, we get a sequence of epochs that matches fairly well with the OJ~287 flare timings. (\citet{LV96}, model 1).

But what generates the optical flares in a Keplerian sequence? While the Keplerian sequence is a purely mathematical rule that gives us the flare times in the historical light curve \citep{D18}, it rather naturally leads to the hypotheses that OJ~287 is a binary black hole (BBH) system. According to the BBH model, a massive secondary black hole (BH) is orbiting a primary BH in an eccentric orbit with 12 year orbital period. The orbital plane of the secondary is not aligned with the accretion disk of the primary BH and the times when the secondary BH crosses the accretion disk will generate a Keplerian sequence. The Keplerian sequence does not depend on the inclination angle between the orbital plane and the accretion disk unless the inclination angle is very small and we take the orbital plane to be perpendicular to the accretion disk. When the secondary BH crosses the accretion disk, it creates an impact on the accretion disk material. A theory of black hole impacts was constructed on general principles of hypersonic impacts on a slab of gas \citep{bon52,P66,H71}. The heated gas escapes perpendicular to the disk on both sides \citep{I98}. The hot bubbles expand and cool down until they become optically thin, and the radiation from the entire volume is seen. This marks the optical flare (\citet{LV96}, model 2). The radiation is unpolarized thermal bremsstrahlung at a temperature of $ \sim 10^5$ K \citep{V12}.

A number of alternative models, for example by \citet{katz97} and \citet{tanaka13}, have tried to explain the single peaks in the old historical light curve of OJ~287, with only one flare per 12 years, or in one case, even with one flare per 24 years \citep{britzen18}. These alternative models go from doppler boosting in the jet, varying accretion rate to precession and nutation of the jet to explain the variability in OJ~287. Though these alternative models explain some features of OJ~287, the BBH model is the most successful to explain the optical flares \citep{D19}. For example, it is one of the major results by \cite{britzen18} that the precession and nutation can explain the flux-density variability in the radio regime. Not only the morphological changes of the jet-structure can be explained but also the variability on different time scales (long-term and short-term). But in optical regime, it is not very successful in explaining the sequence of flares and also the flux rise timescale and low polarization during flares \citep{V08, V16, V17}. In contrast, in the BBH model, the basic orbital period is 12 years which means that there have been five impacts since the BBH model was proposed by \citet{S88}. The root-mean-square error of the predictions of the starting times of the flares has been 16 days \citep{V18}. The model also predicted a number of past flares which were not known at the time, but which have been subsequently discovered from historical plate archives. The total number of flares in the model between 1888 and 2015 is 24, most of which are covered well enough by observations to confirm the model \citep{D18}. In this paper we concentrate on the BBH model which explains the full Keplerian sequence of flare times.

From the orbital period (12 years) and precession rate ($39^{\circ}$ per orbit) of the binary, it is straightforward to calculate the total mass of the system by invoking General Relativity. The total mass turns out to be $\sim 2 \times 10^{10} M_{\odot}$ and the semi-major axis $\sim 0.06$ pc. As far as the parameters of the orbit go, it is not of great importance what actually generates the signal at the crossing points \citep{P98}. But as we mentioned earlier, there is a {\it time delay} between the disk crossing time and the flares and it is important to calculate them to accurately predict the future flare timings and explain the observed flare sequence. The {\it time delay} depend on the properties of the accretion disk.

Since the \citet{LV96} model was calculated, much advance has been made observationally as well as in the theory of Post-Newtonian orbit calculation \citep{D18}. The orbital parameters are derived from a solution with the exact starting  times of 10 well observed flares. Besides the usual orbital parameters, the solution has also two additional parameters which are related to the properties of the standard thin accretion disk \citep{S73}. They determine the time delay between the impact on the disk and the radiation burst. Note that mathematically the problem is strongly overdetermined: 10 flares are needed for a unique determination of the 9 parameters of the problem, but in fact the solution satisfies the timing of all 24 flares during the 130 year interval of the optical light curve. While earlier papers have concentrated on improving the elements of the binary orbit \citep{V11}, in this paper we study the properties of the accretion disk. In particular, we ask the question if the disk model is still self-consistent after a number of changes have been made in the orbital parameters. The self-consistency is judged by the parameters of the standard accretion disk, $\alpha$ and $\dot m$ that follow from the orbit solution.

The time delay for a particular flare dpepends on the accretion disk properties at the impact site, the relative velocity of the secondary BH during impact and very weakly on the relative inclination angle $i$ between the orbital plane and the disk plane via some power of $\delta=\sqrt{1-\cos i}$. The $\delta$ factors are included in the formulae of \citet{P16}, but they are of no importance unless the two planes are far from perpendicular. In the following we assume that $i\sim\pi/2$. Whether the secondary passes through the jet depends on the orientation of the jet which precesses in about 1400 yr cycle \citep{V11}. It is actually quite unlikely that the jet crossing would have happened during the 130 yr period of our observational record, and it is possible that it will never happen in the present configuration of the OJ~287 system.

We make use of the time delays solved in \citet{D18} as they relate directly to the properties of the accretion disk. We update the accretion disk model of \citet{LV96} as well as look in more detail at the various stages of expansion of the radiating bubble. We study the two latest flares in particular where the black hole impact is almost perpendicular to the disk plane: the apocenter flare in 2015 and the pericenter flare in 2007 (and in 2019). The latest of the predicted flares began on November 25, 2015, at the exact centenary of the General Relativity theory by Albert Einstein \citep{V16}. In earlier papers, it was referred to as the GR flare. Here we simply call it the 2015 flare. In the model the flares arise from the impacts of the secondary black hole of mass $1.5\times10^{8} M_\odot$ on the accretion disk of the primary of $1.835\times 10^{10} M_\odot$. The apocenter distance is $(1+e)/(1-e)\sim 5.7$ times greater than the pericenter distance, and by Kepler's second law the impact speeds go in the same (inverse) ratio. The size of the bubble of hot gas extracted from the disk is bigger when the speed is lower, by the rules already worked out by Bondi \citep{bon52}, and consequently the radiating bubble becomes optically thin later, while also the radiating volume is greater. The details have been outlined by \cite{P16}.

The next predicted flare will peak on July 30, 2019 and it is expected to be nearly identical to the well covered 2007 flare \citep{V08}. The 2007 flare allowed the determination of the energy loss to gravitational waves from the OJ~287 black hole binary system while the 2015 flare was used to test the high order Post-Newtonian terms of gravitational radiation at a length scale which has not been accessible to us before \citep{D18,D19}. The detailed study of the 2015 flare light curve is important for understanding the future flares. Sometimes it may not be possible to observe the flares in optical region, which is the likely case for the 2019 flare, due to the closeness of OJ~287 to the sun in the sky. If we manage to obtain observations at other wavebands, then we have to understand the wavelength dependence of the flare timings. We also note that the big flares, which typically arrive in pairs separated by one to four years (and in triples every 60 years), are themselves composed of a double flare, with the component separation of about a week or less. These doubles have different polarization properties. We will present a preliminary theory for the origin of this internal flare structure.

In section~\ref{sec:disk} we describe the disk model used to calculate the time delay and how we update the disk parameters. The shape of the light curve during the 2015 first flare are discussed in sections~\ref{sec:fluxrise} and \ref{sec:fluxdecline} while in section~\ref{sec:second_flare} we concentrate on the second flare. In section~\ref{sec:multi_color} we talk about the multicolor data during 2015 flare and section~\ref{sec:conclude} draws the conclusions.

\section{Disk model and parameters} \label{sec:disk}

 The 1995 \citet{LV96} model was based on a variant of the \cite{S73} theory of thin accretion disks, with the viscosity parameter $\alpha=1$ and the accretion rate $\dot m=0.1$ in Eddington units. This model is based on steady-state accretion of matter toward the central body, in a way that satisfies the basic conservation laws \citep{Sha83}. Later work has demonstrated that this theory is very much valid even today. The main improvement to the theory has been the adding of dynamically significant magnetic fields in the disk. For example \cite{P03}, \cite{B07} and \cite{J19} have demonstrated that $\alpha$ disk models are also possible in the presence of very strong magnetic fields. In fact, it is necessary to introduce these fields to provide stability to the inner accretion disk \citep{S81,S84}. Differences arise based on the assumed strength of the magnetic fields in the disk \citep{P03}. In the most extreme case the magnetic pressure dominates over radiation and gas pressures \citep{J19}. We stay with the more moderate model where the magnetic pressure equals the gas pressure in the disk \citep{S81,S84}. \citet{G79}, \citet{T79} and \citet{C81} have argued that when this threshold value of magnetic flux is exceeded, flux escape will happen in a short time scale compared with other disk time scales.

In this paper, we use thermal equilibrium model of a magnetized accretion disk where the thermal gas pressure and the magnetic pressure are in equilibrium at all disk radii while radiation pressure dominates over both of them by orders of magnitude in the inner disk \citep{S81}. Here we need only the innermost, radiation pressure dominated, region of this model. The orbit of the secondary covers the range from 8 to 60 Schwarzschild radii of the primary black hole which fits inside the innermost region. The disk properties are scaled to our primary black hole mass by the scaling laws of \citet{S84}, and the same laws are later used to adjust the parameters $\alpha$ (denoted usually by $\alpha_g$ in this context; here we leave out the subscript) and $\dot m$. The numerical values of the disk model are given in Table 2 of \citet{LV96}. They do not include azimuthal variations in the disk. Once an accurate orbit model has been worked out, one could in principle include the effects of all past impacts in the disk structure, but it is a complication which has not been attempted so far, and it may not be even within calculation resources at this time.
Using this model, \citet{D18} found tentatively that both $\alpha$ and $\dot m $ should be lower than in the \citet{LV96} model, but the exact values were left uncertain. We first update the \citet{LV96} disk model and then find the disk parameters using time delays calculated in \citet{D18}.

\begin{figure*}
 \centering
 \includegraphics[width=0.45\textwidth]{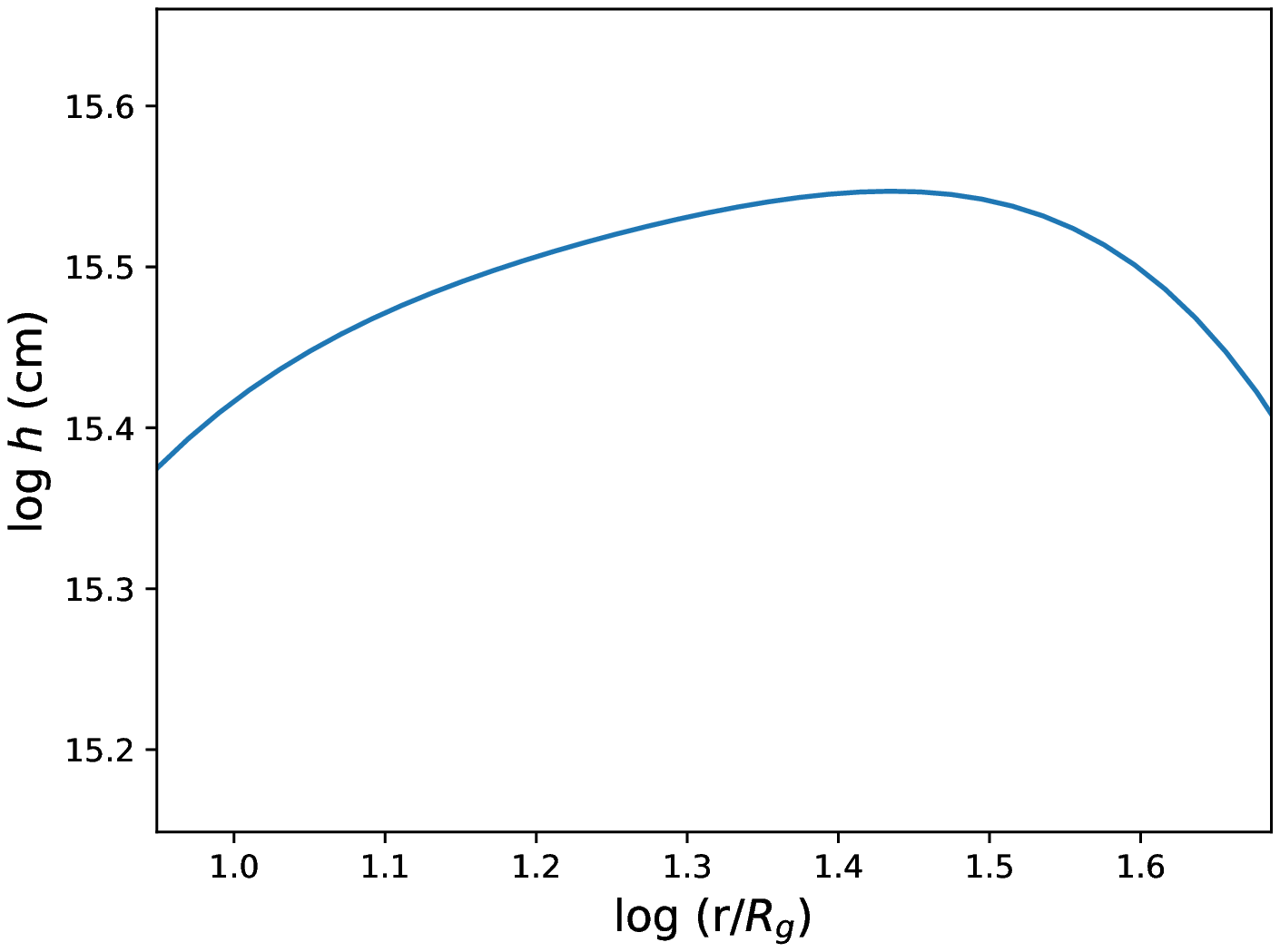}
 \includegraphics[width=0.45\textwidth]{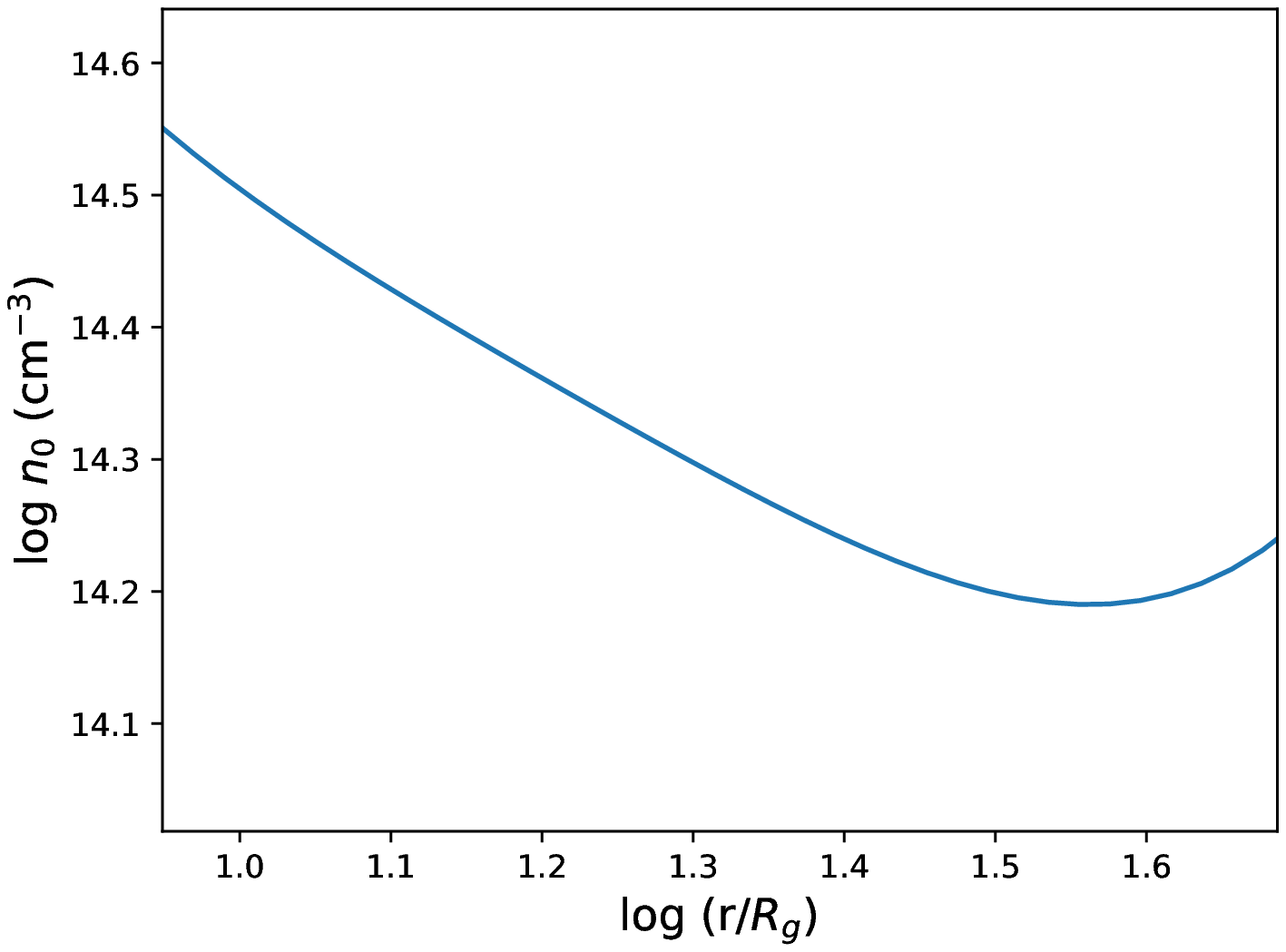} \\
 \includegraphics[width=0.45\textwidth]{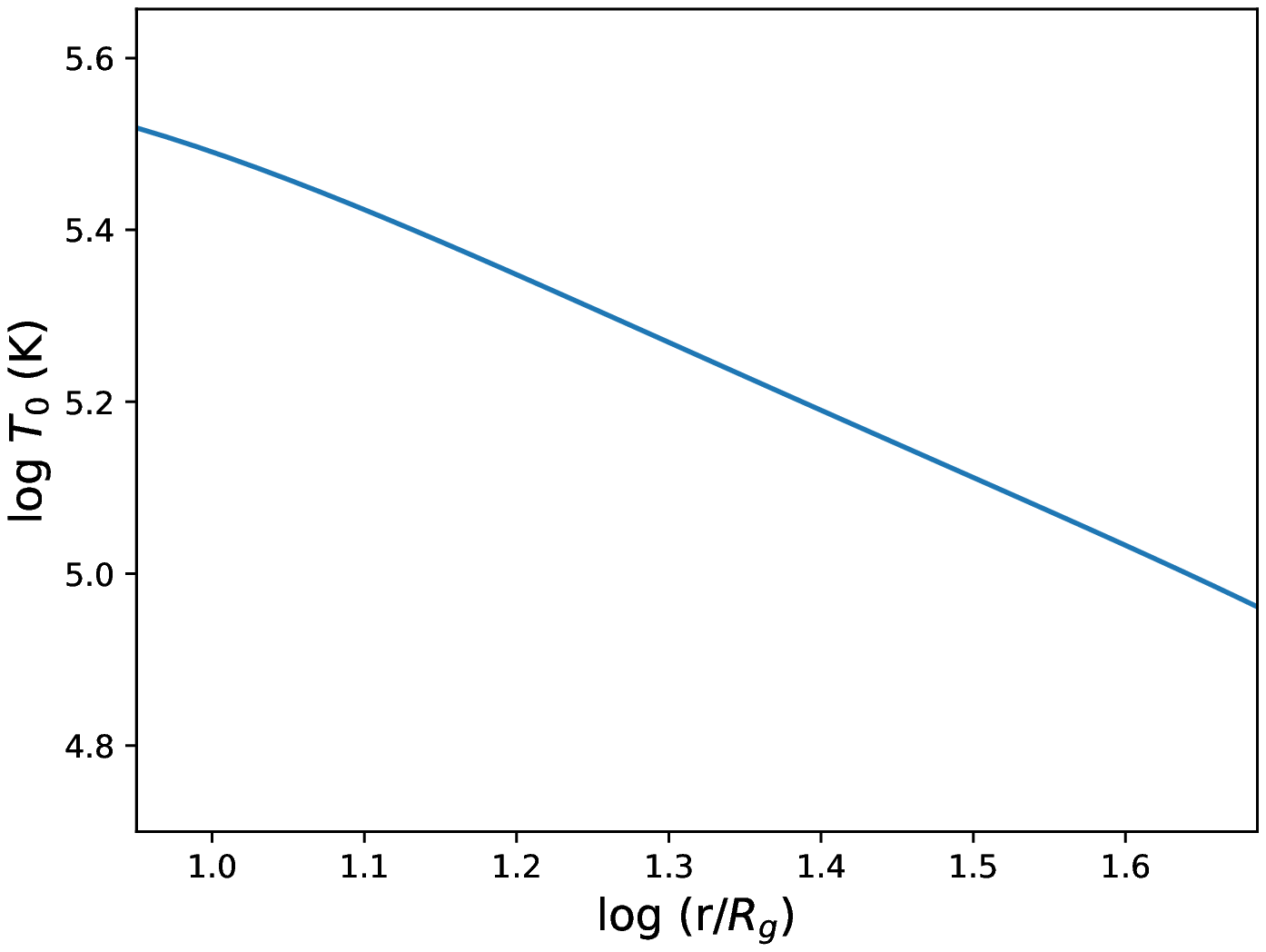}
 \includegraphics[width=0.45\textwidth]{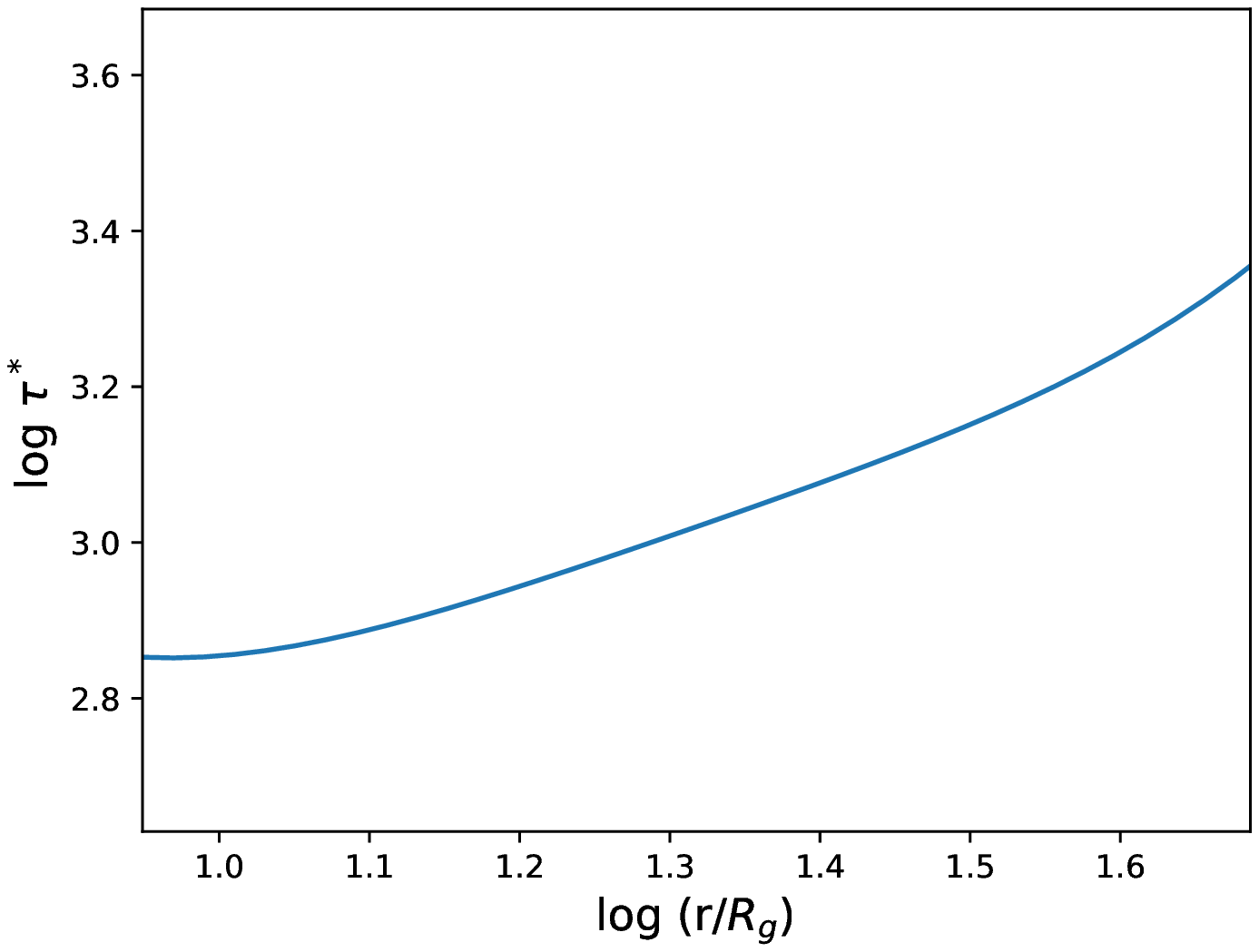}

 \caption{The figure shows various accretion disk properties with updated distance scale. The upper left and right panels show the disk semi-height and number density of particles, respectively. The lower left and right panels show the disk temperature and optical depth, respectively.}
    \label{fig:disk_properties}
\end{figure*}

 In our  Figure~\ref{fig:disk_properties} we first of all correct the distance scale, the radial distance
 from the center of the primary black hole to the position in the disk. \citet{LV96}
 used the primary mass value $1.66\times10^{10} M_\odot$ while the orbit model now
 gives $1.835\times 10^{10} M_\odot$. Since the accretion disk properties are calculated
 for distances with respect to the Schwarzschild radius of the primary $R_g$
 ($R_g = 362~AU$ in the present model) this means a ten percent increase in
 the distance scale. Secondly, \citet{D18} found that the disk thickness should be scaled
 down by a factor of 0.9. The half-thickness $h$ is practically constant (about
 $170~AU$) over the distance range considered here. We adjust the particle density
 $n_0$ upward from \citet{LV96} by a factor of {\bf{$\sim2$}}, to values a little above $10^{14}
 cm^{-3}$, for reasons that are explained later.

 An important quantity is the optical depth $\tau^*$ of the accretion disk at
 different distances from the center. In Figure~\ref{fig:disk_properties} we have adjusted the $\tau^*$
 values to the changes in $h$ and $n_0$, keeping the temperature $T_0$ as
 before. We display $\tau^*$ for one half of the disk height, so that it
 corresponds to the disk semi-height $h$. The impact which generates the 2007 flare
 occurs at the distance of $\sim3200~AU$ which is represented by  the left margin
  in Figure~\ref{fig:disk_properties}, while the properties for the 2015 flare at the distance of 
 $R \sim 17500~AU$ are read from the right margin in Figure~\ref{fig:disk_properties}. These two flares occur
 at the pericenter and the apocenter of the binary orbit, respectively.
 
After the secondary has impacted on the disk, a roughly cylindrical slab of the disk matter is removed from it. The radius of the cylinder  $R_a$ is proportional to the Bondi-Hoyle accretion radius $R_{BH} = \eta R_{sec}$, where $\eta = (c/v_{rel})^2$ and $R_{sec}$ is the Schwarzschild radius of the secondary black hole, $v_{rel}$ is the impact speed and $c$ is the speed of light. In the standard \cite{S73} accretion disk model the inner part of the disk is radiation pressure dominated, and therefore the adiabatic constant $\gamma=4/3$. The density increase in a strong shock is by a factor $(\gamma + 1)/(\gamma - 1)$, and therefore the compression factor is 7. The resulting post-shock sound speed is $\sqrt{8/7} v_{rel}$ where $v_{rel}$ is the speed of impact of the secondary on the accretion disk \citep{L99}. The impact of the secondary BH on the accretion disk influences the disk matter roughly out to the Bondi radius of the impact point \citep{I98}. The Bondi radius is a rather robust quantity, as magnetohydrodynamic simulations have shown \citep{C12}.

We define $C_{BH}$ as the coefficient of the accretion radius $R_a$ in the equation
\begin{equation}
R_a = C_{BH}~ R_{BH}.
\end{equation}
For the case  $R_{BH}/H = 1/3$ the \citet{I98} calculation gives $C_{BH} \sim 0.45$ while for the case $R_{BH}/H = 3$ they find $C_{BH} \sim 0.35$. Here $H=2h$ is the disk thickness. In this work we also cover a range of 9 units in the $R_{BH}/H$ parameter.

The impact velocity on the disk $v_{rel}$  is calculated from $v_{rel}^2 = v_{orb}^2 + v_K^2$, where the orbital speed of the secondary is $v_{orb}$ and the Keplerian speed of the disk is $v_K$. The height of the cylinder may be taken as $h/7 $. Therefore the volume of the cylinder of matter is
\begin{equation}
V_0 = 4\pi/3~ R_0^3 =  \pi~ C_{BH}^2~ (c/v_{rel})^4~ R_{sec}^2~ (h/7).
\end{equation}
while the sphere of radius $R_0$ is defined such that it has the same volume
$V_0$. The Schwarzschild radius of the secondary is $R_{sec} = 3~AU $, and 
$h/7 \sim 24~AU$. For the 2007 flare, $v_{rel} \sim 0.35~c$ and therefore the radius of
the equivalent sphere $R_0\sim  13~AU = 0.0002\,\text{light yr}$. For the 2015 flare, the
corresponding values are $v_{rel} \sim 0.12~c$ and $R_0 \sim 47~AU= 0.00076~ \text{light yr}$. In general, we may write
\begin{equation}
R_0 = 3.73~ (c/v_{rel})^{1.19}~ [h/(170~AU)]^{1/3}~AU.
\end{equation}

The equilibrium temperature $T_{eq}$ of this sphere at different distances from the primary is calculated from
\begin{equation}
 T_{eq}^4 = (18~ n_0~ m_p~ v_{rel}^2)/(7 a),
\end{equation}
where $m_p$ is the proton mass and $a$ is the radiation constant. For the parameters of the 2007 flare, we have $T_{eq} = 2.2\times 10^6$ K, $\sim 6.5$ times higher than the disk temperature. In the 2015 flare we have $T_{eq}=1.06\times 10^6$ K, about 11.5 times higher than the corresponding disk temperature.

For the calculation of the optical depth we use the geometrical mean of the Thomson opacity $\kappa_T$ and the bremsstrahlung plus bound-free opacity. The latter is calculated from
\begin{equation} 
\kappa_a = 6.8\times 10^{22}~ \kappa~ \rho~ T^{7/2} g~cm^{-2},
\end{equation}
where $\rho$ is the matter density and $T$ is the temperature, and $\kappa$ is
\begin{equation} 
\kappa = 1 + 10^3~Z/t_{bf}.
\end{equation}
The first term comes from free-free transitions and the second term from bound-free transitions. $Z$ is the fraction of heavy elements in the interstellar medium, $Z\sim0.02$ and $t_{bf}$ is a coefficient for bound-free absorption, $t_{bf}\sim100$ in the spectral region of interest to us.

For high temperatures, $T \sim 10^6$ K, the coefficient $\kappa$ is practically equal to unity since the bound-free contributions are not important. For lower temperatures $T\sim 10^5$ K the bound-free absorption makes a contribution at about $30\%$ level and thus $\kappa \sim 1.3$ \citep{L99}. These two cases refer to  the pericenter and the apocenter, respectively.

  At the impact the density increases by the shock compression factor of seven over the initial density. Comparing with the optical depth in the semi-disk, for the bubble radius $R_0 \sim 13~AU$, $\tau$ is modified by factors relating to the geometrical thickness, density and temperature, and we get
\begin{equation}
 \tau = (13/170) \times7^{3/2} \times 6.5^{-7/4} \tau^* \sim 38. 
\end{equation}
since the initial optical depth of the half-disk $\tau^*\sim 713$ (Table~\ref{tab:PROPERTIES OF THE EXPANDING BUBBLE}). The bubble expands by a factor $\xi=\tau^{4/7} \sim 8$ before the optical depth drops to  $\tau=1$, the threshold for transparency \citep{P16}. This produces the flare. For the GR flare we get similarly $\xi\sim18$. 

 \begin{table}
    \centering
    \caption{The properties of the expanding bubble as a function of distance $r$ from the primary black hole.}
    \label{tab:PROPERTIES OF THE EXPANDING BUBBLE}
    \begin{tabular}{rcccccccr} 
\hline
\multicolumn{1}{c}{r}& $\tau$   & log($T_{eq}$) & $v_{sec}$ & $v_{rel}$ & $R_0$ & $R_{bubble}$ & $t_0$ \\
\multicolumn{1}{c}{(AU)} & & (K)           & (c)       & (c)       & (AU)  & (AU)         & (yr) \\
\hline
~3218 & 38                  & 6.34          & 0.275     & 0.356     & 12.5    & 100          & 0.011      \\
~4027 & 38                  & 6.30          & 0.250     & 0.319     & 14.9    & 119          & 0.020        \\
~5722 & 44                  & 6.23          & 0.207     & 0.264     & 19.6    & 170          & 0.060\\
~7417 & 51                   & 6.18          & 0.174     & 0.226     & 24.2    & 228          & 0.127\\
~9114 & 59                  & 6.14          & 0.150     & 0.198     & 28.5    & 294          & 0.230\\
10809 & 70                  & 6.10          & 0.131     & 0.177     & 32.6    & 368          & 0.391\\
12503 & 83                  & 6.08          & 0.115     & 0.160     & 36.3    & 454          & 0.643 \\
14201 & 101                  & 6.05          & 0.101     & 0.144     & 40.2    & 562          & 1.028 \\
15992 & 126                  & 6.04          & 0.088     & 0.131     & 43.8    & 694          & 1.635\\
17595 & 158                  & 6.03          & 0.064     & 0.120     & 46.7    & 842         & 2.407          \\

        \hline
    \end{tabular}
\end{table}

The main observable parameters are the time delay $t_0$ between the disk impact and the flare,  and the peak flux of the flare $S_V$ as well as the rise time of the flare $t_{rise}$. To simplify matters, we use the formulae (22)-(25) in \citet{P16}, and substitute the appropriate parameter values separately for the pericenter and the apocenter of the orbit. The two cases are obtained by choosing the values of the coefficients A, B and C: A=B=C=1 at the pericenter and A=225, B=2.6 and C=6.5 for the apocenter. We get
\begin{equation}
   t_0 = 0.01\times A~ \kappa^{0.43}~ n_{LV}^{0.91}~ C_{BH}^{0.62} yr 
\end{equation}
\begin{equation}
   S_V = 12\times B~ \kappa^{-0.71}~ n_{LV}^{0.36}~ C_{BH}^{0.52}~ e^{-\psi} mJy
   \label{eqn:peak_flux}
\end{equation}
where 
\begin{equation}
\psi = 0.12\times C~ \kappa^{0.29}~ n_{LV}^{0.36}~ C_{BH}^{0.19}.
\end{equation}
Here $n_{LV}$ is the number density in the disk, in units of the \citet{LV96} value. It is convenient to scale the density value given in \citet{LV96} by this number, as it is the same at all impact distances.

 At the pericenter the observed value of $t_0$ is 0.011 yr \citep{D18}. With $\kappa = A = 1, C_{BH} =0.45$, we find that $n_{LV}=1.94$, i.e., in units of $10^{14}~cm^{-3}$, $n_{14} = 3.2$. For the same parameter values, $\psi \sim 0.17$ and $S_V \sim 9$ mJy. At the apocenter we get for $n_{LV}=1.94$,  $t_0 = 2.4$ yr, if we put $\kappa = 1.3$ and $C_{BH} = 0.35$. Then  $S_V$ becomes $\sim 7.8$ mJy.

We find that we recover the `observables' using the increased value for $n_{14}$ \citep{D18}. The factor of $\sim2$ increase in the density does not seem like much, and if it were not affecting the observed delay times, we would have little hope of detecting it. However, we still need to discuss what are the implications of this change in density. The disk density is a function of the viscosity parameter $\alpha$ and mass accretion rate $\dot m$, here given in Eddington units. Using the scaling relations by \citet{S84} (equation A2) we find that
\begin{equation}
2 h = 6\times 10^{16}~ \dot m~cm = 5\times 10^{15} ~cm,
\end{equation}
where the latter is an `observable' \citep{D18}. Thus $\dot m \sim 0.08$. From equation A5 of \citet{S84} we get at the pericenter
\begin{equation}
n_{14} = 0.4\times \alpha^{-0.8}~ \dot m^{-0.4} = 3.2
\end{equation}
With $\dot m =0.08$, it follows that $\alpha = 0.26$.

These values may be compared with \citet{M11} who find the minimum kinetic power for one of the jets that corresponds to $\dot m = 0.016$. For two jets, and with some of the accretion not ending up in the jets, our value is consistent with theirs. The value of $\alpha$ falls in the range $\alpha=0.23\pm0.1$ estimated by \citet{C81} in a magnetic flux cell model (quoted in \citet{S81}), and agrees with $\alpha\sim0.27$ from steady state models by \citet{S18}. Even though in different scales, also for dwarf nova and X-ray transient outbursts one finds $\alpha=0.25\pm0.15$ \citep{K07}. Our values, with estimated uncertainties, are $\dot m = 0.08\pm0.02$ and $\alpha=0.26\pm0.05$. The uncertainty arises mostly from the uncertainty in the coefficient $C_{BH}$ which has been determined from the illustrations in \citet{I98}. The accretion rate corresponds to about 6$ M_\odot$ per yr.

We may also consider the uncertainty arising from the disk model. If the disk is magnetic pressure dominated, \cite{J19} find that the density $n_{14}\sim 0.6$ and $h\sim 1.1 R_g$ when $\dot m = 0.07$, while $n_{14}\sim 0.9$ and $h\sim 3 R_g$ when $\dot m = 0.22$. \citet{P03} show that such a model is close to the \citet{S73} model with $\alpha=1$, while the surface density in the latter model is independent of the central mass. Therefore we may scale the \cite{J19} values, calculated for central mass $5\times 10^{8} M_\odot$, to our case of $h\sim 0.5 R_g$, and obtain $n_{14}\sim 1.4$ for $\dot m = 0.07$, while we get $n_{14}\sim 6$ when $\dot m = 0.22$. Thus our earlier density value is close to the geometric mean of these two models. The geometric means of $\dot m$ and $\alpha$ are $\sim 0.12$ and $\sim 0.15$ in these models, respectively. In order to cover these possibilities, the range of uncertainty should be widened to $\dot m = 0.08\pm0.04$ and $\alpha=0.26\pm0.1$.

\section{Flux rise of the first flare} \label{sec:fluxrise}

\begin{figure}
 \includegraphics[width=2.4in,angle=270]{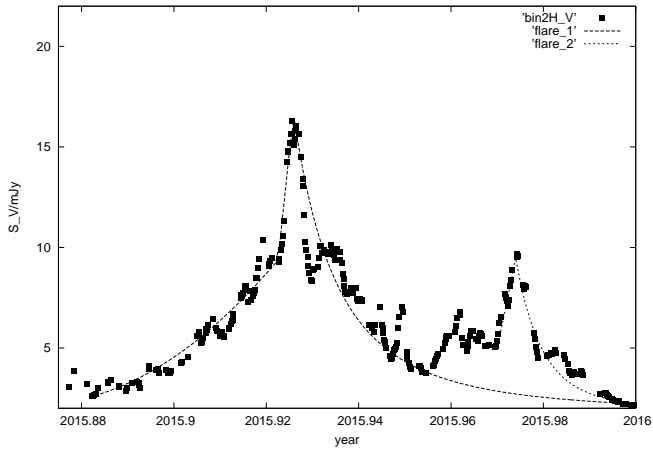} 

    \caption{Observations of the 2015 flare in OJ~287, compared with evolutionary curves. We plot two-hour averages of the R-magnitude, translated to the V-band by V-R = 0.45. The theoretical line is composed of several sections: the flux is proportional to (1) $t^2$; (2)  $t$; (3) $t^{-2.5}$; (4) $t$; and to $t^{-6}$, as explained in more detail in the text.}
    \label{fig:2015_lc}
\end{figure}

The shape of the flare light curve is now calculated from the following principle: during the stage of the flux rise, the expansion of the radiating bubble is relatively slow, starting from the current speed of sound. While the source becomes transparent, the expansion speed rises, and at some point in time, the source becomes fully transparent. It is assumed that the radiation flux is proportional to the volume which is visible to us at each stage. Finally the radiating volume is fully transparent, and the flux rise is due to the visibility front advancing into the source. The peak flux occurs when the whole radiating volume is in our view. Thereafter the bubble cools down and the emission of radiation declines. The rate of decline of the emission depends on the dominant radiation mechanism at each stage. The curves presented in Figures~\ref{fig:2015_lc} and \ref{fig:peaks_comparison} are based on this simple analytic prescription.

The flux rise of the first flare may be divided into two parts {(Figure~\ref{fig:2015_lc})}: a slow
rise from 3 to 10 mJy, and the fast rise from 10 to 16 mJy. We will
discuss the fast flux rise-time $t_{rise}$ first, and the slow rise in section~\ref{sec:fluxdecline}. We may assume that it tells the
size of the radiating region, if the radiation emission front advances into the
source with the speed of light. We take the fastest rising portion of the light
curve for this estimate. From the observations the rise happens in about 0.7
days at the pericenter and about 2.1 days at the apocenter. Considering the
redshift, this corresponds to the radiating bubble size $R_{bubble} \sim 0.5$
light days = $100~AU$ at the pericenter, and three times greater at the
apocenter.

At the apocenter, $R_{bubble} \sim 900~AU$ while we expected about $300~AU$. The
difference is by a factor of $\sim$ three. This suggests that instead of being perfectly spherical, the expanded bubble is
more like a sphere compressed along the line of sight, i.e., along the disk
axis. The compression will reduce $t_{rise}$ by this factor. Let us then consider a source which is a prolate spheroid viewed from the direction of its minor axis.

In the case of the 2015 flare, the extracted matter comes from the disk as a
sheet with the initial diameter-to-thickness ratio of about 10. Apparently it
remains a sheet or a spheroid above the disk even after the expansion, even
though with a lower diameter-to-thickness ratio. 

The expansion could be forced
primarily in the lateral direction by external magnetic pressure from the
corona above the disk. Such a pressure could come from a magnetic corona, as e.g. in the Polish doughnut model by \citet{A78}. The formation of a magnetic corona is a necessary property
of the magnetic disk model \citep{S81}. 

In an adiabatic compression $t_0$ is proportional to $\tau^{6/7}$ \citep{P16}. For a spheroid of compression factor of 3 the optical thickness $\tau$ along the axis of compression is modified by a factor of $1/3 \times 3^{3/2} \times 3^{-7/12} \sim 1$, where the three factors arise from changing geometrical thickness, density and temperature, respectively. Therefore the value of $\tau$ remains practically unchanged, and consequently also $t_0$. 

However, the compression affects the flux. With the factor of 3, the flux is proportional $3^2 \times 3^{-1} \times 3^{-1/6} = 3^{0.83} \sim 2.5$, where the factors relate to increased density, decreased volume, and increased temperature, respectively. Taking into account also the exponential factor $e^{-\psi}$ in equation~\ref{eqn:peak_flux}, this will raise the flux to $S_V\sim 26$ mJy, twice as high as the observed value in the 2015 flare ($S_V\sim13$ mJy). We see that the density factor, entering at the second power, overcomes the volume factor, while the changing temperature plays a smaller role. 

The same calculation may be repeated for an oblate spheroid. A deformation of
a spherical plasma cloud towards an oblate spheroid is expected in an external magnetic field \citep{G54}. The result is the same, if we again view the spheroid along the minor axis, and the spheroid volume is 1/3 of the original spherical volume. 

 The initial temperature of the radiating bubble $T_{eq}\sim10^6$ K at the apocenter and $2.2\times 10^6$ K at the pericenter. At these temperatures the bremsstrahlung spectrum in the optical--UV region is essentially flat, independent of frequency. It means that the expansion timescale $t_0$ is also independent of the frequency, and flares occur simultaneously in optical and UV wavebands, as was observed by \citet{V16}.

This is important for the pericenter flare in 2019: if observed with the Spitzer Space Telescope \citep{L18}, it will produce the infrared flares at the same time as in optical. This is because at the high temperature of the 2019 flare $\kappa=1$ in both wavelength regions to a high level of accuracy.

The compression reduces the effective expansion factor and will therefore affect the temperature of the radiating bubble. The effective temperature becomes about $8\times 10^4$ K for the 2015 flare and $3\times 10^5$ K for the 2019 flare. Since the 2015 flare was observed also in ultraviolet by \citet{V16}, we may use the flux ratio between the optical and UV bands to confirm the temperature. Taking  $\kappa \sim 1.25$ times greater in ultraviolet than in optical \citep{L99}, and assuming that the extinction in the OJ~287 host galaxy is 0.15 magnitudes greater in ultraviolet than in optical \citep{V12,G95}, we derive the above temperature. That is, we may derive the temperature $T\sim8\times 10^4$ K directly by using observations.

\section{Flux decline of the first flare} \label{sec:fluxdecline}

\citet{I98} show that outside the bubble where photons are trapped, a high velocity outflow develops which they call fountains. In the fountains the outflow speed rises to about twice the relative impact velocity. After the optical transparency, the photons are free to move out and the whole bubble becomes part of the expanding fountain. Therefore, we will consider the evolution of the brightness of the bubble at this stage. The fountains expand into the corona where the sound speed is high due interactions of the tenuous coronal gas with the fast secondary \citep{D18}. The coronal sound speed determined from the transmission of particles from the impact site to the jet via the corona has been estimated as $\sim 0.22~c$ \citep{V17}. This could be effectively the expansion speed of the bubble in its transparent stage.

The slow rise time in the first flare occurs at a rate which corresponds to the speed of sound, initially $\sim v_{rel}/4$ and then rising uniformly to $v_{rel}$ \citep{I98}. The line in Figure 1 is based on this rule if the flux is directly proportional to the increasing volume of the transparent source. This is the rate at which we expect the bubble to become transparent, if there is a density gradient in the bubble, with the density increasing inwards. When the whole bubble has become transparent, the rest of it enters our view with the speed of light $c$. This transition surface seems to divide the bubble roughly in two. If it were a spheroid, then the whole spheroid becomes transparent at the time when our increasing viewing volume has reached the center of the spheroid. 

The bremsstrahlung luminosity of the bubble evolves as $\sim(R/R_{bubble})^{-5/2}$ where the radius $R$ may be taken as $R \sim 2 v_{rel}\times t$ and $t$ is the time since the optical transparency $t_0$. Consequently, the brightness evolves with time as $(t/t_0)^{-5/2}$.

\section{The second flare}
\label{sec:second_flare}

In the 2015 flare, as in other earlier flares, the unpolarized thermal flare is followed by a highly polarized flare. \citet{S85} observed the degree of polarization of the first flare in 1983 to be close to zero while they found the second flare of the 1984 outbursts to be highly polarized. \citet{V10} reported a high degree of polarization in the second flare in 2005, while \citet{V08,V16} measured the polarization in both the first and the second flares in 2007 and 2015. They all follow this general rule, with the first flare being essentially unpolarized and the second flare reaching the $40\%$ level of polarization in some cases.
  
The high degree of polarization is probably an indication of the synchrotron nature of the radiation in the second flare. The presence of synchrotron flares implies strong magnetic fields in parts of the source volume. These as well as highly relativistic electrons arise naturally in the strong shocks of the black hole - accretion disk collisions \citep{M99,G07}.
 
The regions with higher opacity due to synchrotron absorption are embedded in the rapidly expanding bubble, and they become optically thin later and produce their own flare. Since the bremsstrahlung flare intensity is only about 1/2 of the expected value in the 2015 flare, we may assume that around 1/2 of the source volume is occupied by matter where synchrotron emissivity dominates. The synchrotron flare peaks at the time when the bremsstrahlung flare has declined by a factor of about 20, that is the bubble has expanded by another factor of $\sim3$. At that time the synchrotron flux is about 10 times higher than the bremsstrahlung flux. If the radiating volumes are similar, the synchrotron emissivity is about an order of magnitude greater than the bremsstrahlung emissivity.

The underlying jet spectrum of OJ~287 is well described by a synchrotron self-Compton model for suitably chosen parameters \citep{C07,V12}. In contrast, the spectrum from the secondary flare does not follow such models. This is most clearly seen from the x-ray flux during the second flare: it does not follow the optical flux, as would be expected if it were coming from the jet \citep{V16}. The spectrum of this flare from optical to ultraviolet is  $\propto\nu^{-\mu}$ where $\mu\sim 1$ ($\nu$ is the frequency). This is the expected spectrum for a synchrotron source for a spectral region modified by energy losses. The jet emission comes from a highly beamed source of a large Lorentz factor, while the expanding bubble is only mildly relativistic. Also the mechanisms for generating the jet flares and the bubble flares could be quite different. Therefore it is not surprising that the syncrotron flare is different from flares arising in the jet.

\begin{figure}
 \includegraphics[width=2.4in,angle=270]{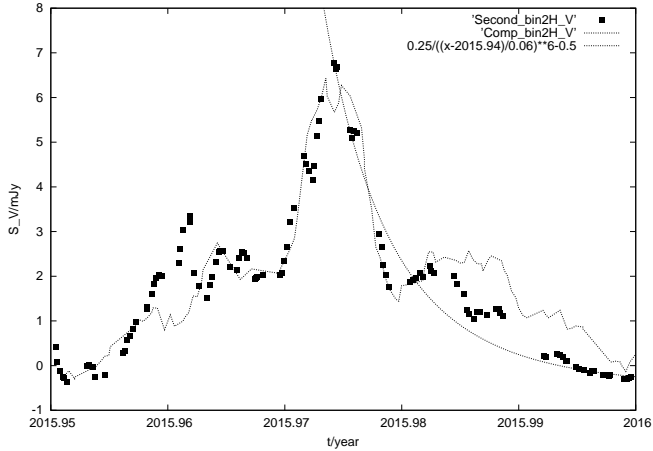} 

    \caption{Comparison of the two 2015 flares in OJ~287. The observations of the second flare are plotted by squares while the modified light curve of the first flare is given by the non-smooth line. The modification of the first flare is done by stretching the time scale by a factor of 1.43, scaling down the flux range by a factor of two, and advancing the time to match with the time of the second flare. The smooth line shows the flux decay as $flux\sim t^{-6}$.}
    \label{fig:peaks_comparison}
\end{figure}

The light curve of the second flare is almost like a slow-down repeat of the first flare, with a smaller amplitude. We compare the two flares in Figure~\ref{fig:peaks_comparison}. The base level of the first flare has been deducted from the second flare, and the observations are plotted by squares. The comparison curve is for the first flare when a slow-down factor of 1.43 has been applied and the size of the flare has been reduced by a factor of two. 

We may assume that the flux rise again gives the size of the radiating region. The rise time is now about 1.43 times longer than in the bremsstrahlung flare. In the bremsstrahlung flare this flux rise represents roughly one half of the source depth, i.e., for the synchrotron source the source depth is $2\times1.43\sim3$ times bigger if the visibility front again advances into the source with the speed of light. The flux decreases as $R^{-2(2\mu+1)}$ or $R^{-6}$ in a synchrotron source of $\mu = 1$ \citep{S60}, which is steeper than in the first flare. We see this difference clearly at the tail end of the flares in  Figure~\ref{fig:peaks_comparison}.

\section{Multicolor light curves}
\label{sec:multi_color}

  We have been monitoring OJ287 brightness at the Cracow (KRK) and Mt. Suhora (SUH) observatories since 2006/7 season. We gather measurements in the R wide band filter, occasionally also in UBVI filters. In order to achieve a dense coverage of the predicted 2015 flare, a multisite campaign was organized and most sites observed the target  in the R filter. The observers provided raw images and their calibration for bias, dark and flatfield was performed with the $IRAF$ package. To secure as uniform results as possible, the magnitudes were extracted using the $CMunipack$ software (an interface for the $DAOPHOT$ package) by a single person and using the aperture method. The same comparison and check stars were used for each site and throughout the entire campaign. We combined data from different observatories based on the measurements of comparison and check stars magnitude differences. The data discrepant by more than 6~$\sigma$ were simply discarded and the resulting light curve was left in the instrumental system. In the 2015/16 and 2016/17 seasons we collected more than 18 thousand single points in the R filter alone. The resulting mean magnitude difference between comparison and check stars is -0.584 $\pm$ 0.002 mag (formal uncertainty).
 Further details about the participated sites can be found in \cite{V16}. The light curve of OJ287 has a dense coverage in the R filter but observations in other bands are scarce. Therefore we appended the Cracow and Mt. Suhora data with observations taken with the Nordic Optical Telescope (NOT) and with those published by \citet{G17}.
 
 \begin{figure}
 \includegraphics[width=2.4in,angle=270]{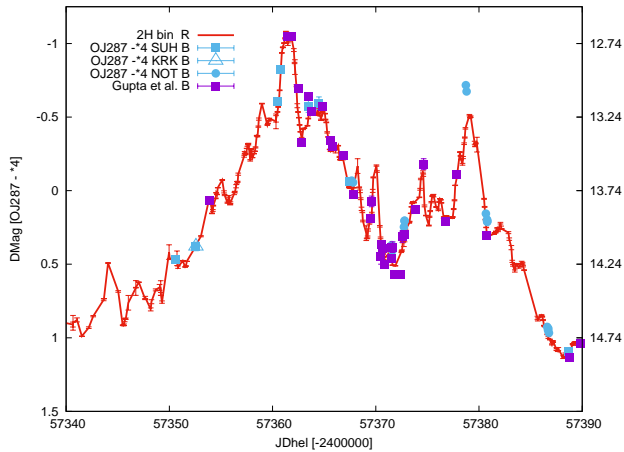} 

    \caption{The 2015 flares in OJ~287 in two colours: R-band (red line with crosses) and B-band (blue and violet symbols). The red line has been raised so that it matches the blue symbols at the peak of the first flare. There is no evidence for a timing difference between the two colours. The second peak is exceptionally blue. The result appears significant even though it is based only on two measurements.}
    \label{fig:R_B_2015lc}
\end{figure}

  Figure~\ref{fig:R_B_2015lc} compares the R-band and B-band light curves, with an adjustment for the R-band data to match the two bands at the peak of the flare. The R-band light curve is from \cite{V16}, binned at two-hour intervals. There is obviously no timing difference between the two colours. There may be a slight blue-excess in the second peak. Unfortunately this peak was not well covered in the multiband data.

\section{Conclusions} \label{sec:conclude}

We have updated the accretion disk parameters in the OJ~287 accretion disk so that they agree with recent observations and theoretical orbit models. We find that the effective viscosity parameter $\alpha=0.26\pm0.1$ rather than unity, as was previously assumed \citep{LV96}. This value agrees with theoretical expectations for a magnetic disk, and to this extent we have demonstrated that our model is self-consistent. It does not prove that it is the only possible model, but we have argued that even if the magnetic flux is much greater than in our model, the disk parameters should still stay within our error limits. The value of the accretion rate, $\dot m=0.08\pm0.04$, puts the model out of the range of models like the slim disks or ADAF. We also find that the thermal flares should occur simultaneously in the spectral region from infrared to ultraviolet. This is important in cases where we cannot cover the whole spectral range with observations, as is likely with the next thermal flare on July 30, 2019, due to closeness of OJ~287 to the Sun. So far only IR observations with Spitzer have been scheduled \citep{L18}. The exact time of the flare in optical (that is, the hour of the flux peak on July 30) may be used to test the no-hair theorem of black holes \citep{D18}.

The simultaneity of flares from infrared to ultraviolet is a property of bremsstrahlung flares. It does not hold for flares that have a black body spectrum \citep{P16}. This will be tested in the coming Spitzer observations.

The radiating bubble appears compressed along the disk axis by about a factor of three in the 2015 flare. Also the flares seem to have two distinct radiation regions, one where the thermal bremsstrahlung dominates, and the second one where synchrotron emissivity is more important. The first region becomes optically thin earlier and produces an unpolarised flare, while the second region gives rise to a highly polarised flare later on. This appears to be the general light curve feature at all impact distances. The main difference is in the time scales: pericenter flares are faster by a factor of three since the geometrical thickness of the radiating volume varies by this factor.


\acknowledgments
MV acknowledges a grant from the Finnish Society of Sciences and Letters, and SZ acknowledges the NCN grant number 2018/29/B/ST9/01793.

%



\end{document}